\documentclass{article}
\usepackage{graphicx}
\begin{document}

\title{Very high quality factor measured in annealed fused silica}
\author{Alexandr Ageev$^1$, Belkis Cabrera Palmer$^1$, Antonio De Felice$^1$, \\
		Steven D. Penn$^{1,2}$, Peter R. Saulson$^1$ \thanks{E-mail address: saulson@physics.syr.edu}\\
 	{\small\sl $^1$Department of Physics, Syracuse University, Syracuse, NY 13244-1130,}\\ 
	{\small\sl $^2$Department of Physics, Hobart and William Smith Colleges, Geneva, NY 14456}}

\maketitle

\begin{abstract}

We present the results of quality factor measurements for rod samples made 
of fused silica. 
To decrease the dissipation we annealed our samples. 
The highest quality factor that we observed was $Q=(2.03\pm0.01)\times10^8$ 
for a mode at 384~Hz. This is the highest published value of $Q$ 
in fused silica measured to date.

\end{abstract}

\section{Introduction}

New large interferometric detectors of gravitational waves, 
such as LIGO~\cite{LIGO}, VIRGO~\cite{VIRGO}, 
GEO~600~\cite{GEO} and TAMA~\cite{TAMA} are now  coming on line. 
To reach their design sensitivities, many noise sources will need to be understood and controlled. 
One of the most fundamental noise sources  that  
limits sensitivity is internal vibration of the test masses due to thermal noise. 
To reduce the thermal noise resulting from the test masses, these masses must be made from low loss material and must have low dissipation in the fundamental modes of the mass and its suspension.  
There are many dissipation processes in materials that could limit the quality factor; 
including thermoelastic dissipation, surface dissipation and dissipation 
due to lattice defects\cite{Zener, VBthermal}.
For fused silica fibers with diameter less than a few mm, the dissipation in the surface is much more important than
losses in the bulk material; in this regime the dissipation decreases with 
increasing the diameter of the fiber~\cite{Andri}. 
The surface loss in fused silica has not been fully explained by any single
loss mechanism, although hypotheses have been proposed for several
contributing processes, including adsorbed water \cite{Mitr},  surface
bond oscillations \cite{Bartenev}, microcracking \cite{Lunin}, and absorbed
alkali molecules\cite{Zdaniewski,  Fraser}.
A vacuum annealing treatment was shown by Numata and collaborators to improve the
quality factors of fused silica for all resonance frequencies~\cite{Jap}. Annealing in air also improves 
the quality factor, but can cause surface defects and increase surface losses~\cite{Lunin}.


In the work presented here, we measured the quality factor of the transverse vibrations of fused silica rods, 
with large diameters (up to 12 mm) to reduce dissipation from the surface. 
We were especially interested in exploring the effects of an annealing treatment, which we performed in an argon atmosphere.

\section{Experiment}

\begin{figure}
\centering
\includegraphics {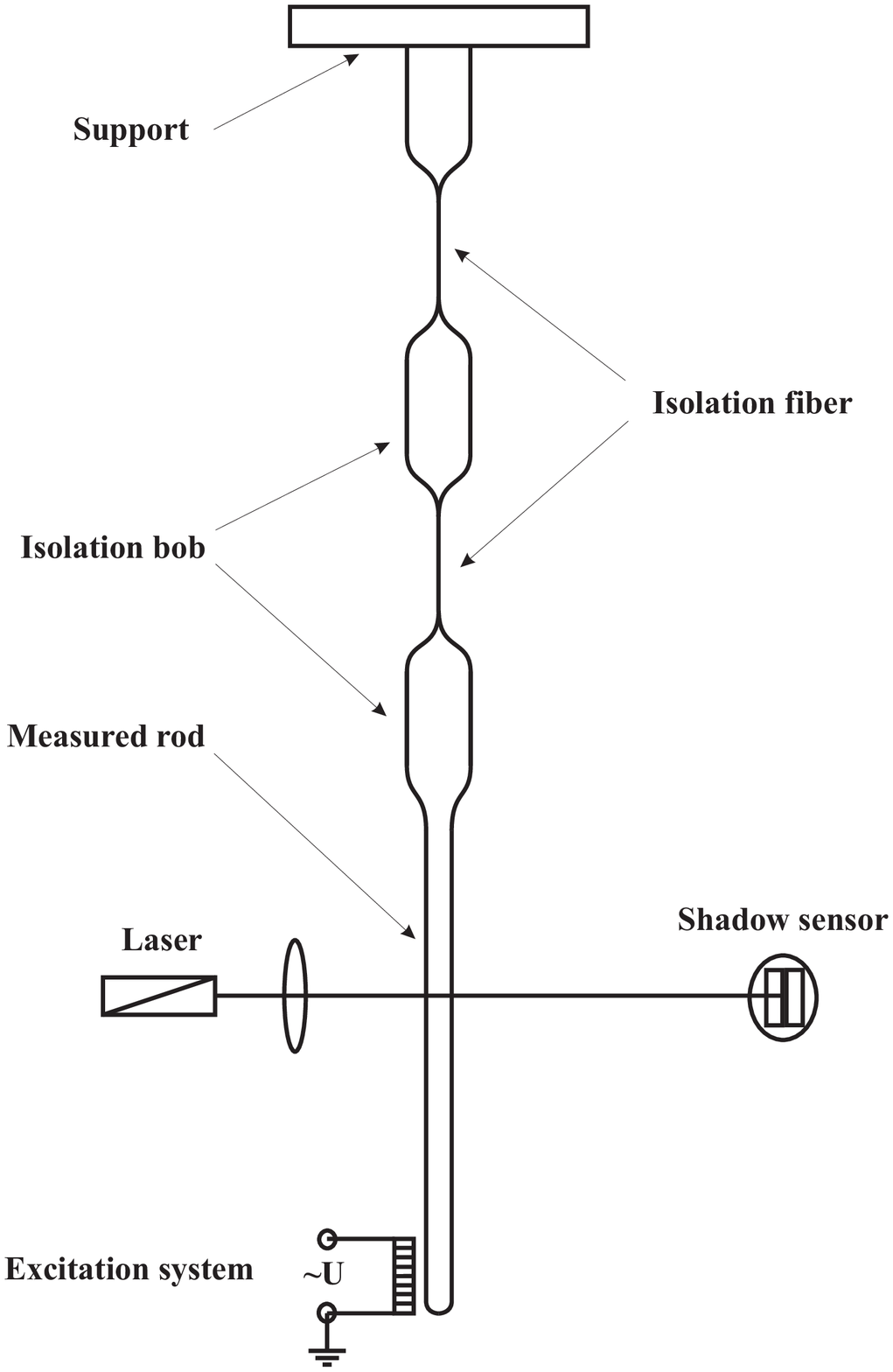}
\caption{Experimental setup}
\label{setup}
\end{figure}

The measurement method was the same as with previous experiments performed in our laboratory at Syracuse 
University~\cite{Andri, Steve}.  
The quality factor of the rod was measured using the ringdown method.
The sample was suspended in a vacuum chamber by an all silica isolation chain of fine fibers and 
massive isolation bobs, as shown in Fig.~\ref{setup}.

Our samples were made of Suprasil~2, Suprasil~312 and  Suprasil~312~SV  fused silica 
from Heraeus Amersil, Inc. The material was supplied as cylindrical rods about 20 mm in diameter.
Each of our samples was made in the Syracuse University glass shop using a glass lathe and 
a natural gas flame.
The section of the sample that was to be measured was drawn to the required diameter and length from the original material.
The length of the measurement rod for all samples was between 24 and 29~cm; diameters were  
3~mm, 8~mm or 12~mm. 

The end result of the fabrication process was a structure made from one piece of material, 
consisting of the measurement rod,  an isolation bob 10~cm in length and 20~mm in diameter and 
a smoothly tapering section between these two rods $2$ to $3$~cm in length. 
The process of preparing the sample took about 2~hours. 

The subsequent annealing treatment was performed in a continuous flow of high purity argon gas 
to remove any expelled impurities and prevent the adsorption of impurities.
Samples were heat treated for 10~days: 3~days were devoted to slowly heating the sample, 
which was then held at the annealing temperature of  950~$^\circ$C for 1~day. 
Finally, the sample was cooled over 6~days to room temperature. 
We chose the long annealing period and long cooling period in order to decrease internal density fluctuations 
that could be responsible for additional dissipation.
As shown in papers~\cite{Parc, Sakaguchi}, annealed silica samples have decreased levels of density fluctuations, 
as shown by a lower level of Rayleigh  scattering. The Rayleigh  scattering intensity is also lower 
in slowly cooled samples, for a given annealing temperature.

Flexural modes of the measurement rod were excited by the gradient of an applied electric field oscillating at the resonance frequency. 
The source of electric field was a comb capacitor placed near the end of the rod. After excitation the capacitor was grounded and the mode was allowed to ring down. 

We used the shadow method for measurement of the vibration of the sample. 
A light beam from a diode laser cast a shadow of the sample on a split photodiode.
The photocurrents from the two halves of the detector were fed through a differential amplifier, whose
output in turn went to a bandpass filter,
a lock-in amplifier and computer data acquisition system. The signal of the detector contains  
the spectral component corresponding to the oscillation of the sample. 
For extraction of the amplitude of the oscillation, we applied a reference sinusoidal  signal 
to the lock-in amplifier, with frequency slightly offset from that of the resonance. 
Hence, our data was a damped 
time series  with frequency $\bigtriangleup f=f_{\mathrm res}-f_{\mathrm ref}\approx 0.3$ Hz. 
The amplitude decay time $\tau$ was fitted from the recorded time series. The quality factor was calculated 
using the  formula $Q=\pi f\tau $,   
where $f$ is the mode frequency. The time of each measurement was about 5~$\tau$, ranging for different modes from 3 hours to 8 days. 
All measurements were made at room temperature, at a working pressure of $10^{-6}$~Torr.

For each sample, we measured the quality factor of three or four modes.
The quality factor of each mode was measured three times; repeated results had a difference not higher than~$5\%$. 
The statistical error of the fits was less than~$1\%$.

\section{Results}

\begin{table}[tbp]

   \begin{center}  
      \begin{tabular}{ccccc} 
      \hline  

       Brand          & Diameter, mm 	 & $Q^{-1}_{\mathrm initial}$ 			& $Q^{-1}_{\mathrm annealed}$    \\

      \hline  

      Suprasil 	    & 0.17	  		 & $16.1\times 10^{-8}$ 	& $14.7\times 10^{-8}$ \\
      2	          & 6                  & $6.25\times 10^{-8}$		& 			     \\
                      & 8                  & $5.32\times 10^{-8}$		& 			     \\
                      & 12                 & $4.35\times 10^{-8}$		& 			     \\ 
      \hline  

	Suprasil        & 3	             & $1.37\times 10^{-8}$ 	& $1.00\times 10^{-8}$ \\
      312		    & 8 			 & $1.22\times 10^{-8}$ 	& $0.49\times 10^{-8}$ \\
                      & 12 			 & $1.22\times 10^{-8}$ 	& 			     \\
                     
      \hline  

      Suprasil 	    & 8  			 & $3.28\times 10^{-8}$ 	& $1.89\times 10^{-8}$ \\
      312 SV	    & 12 			 & $3.10\times 10^{-8}$ 	& $1.64\times 10^{-8}$ \\
			  	
     \hline  

  \end{tabular}
  \end{center}
\caption{Dissipation of fused silica samples before and after annealing.}
\label{table1}
\end{table}

\begin{figure}
\centering
\includegraphics {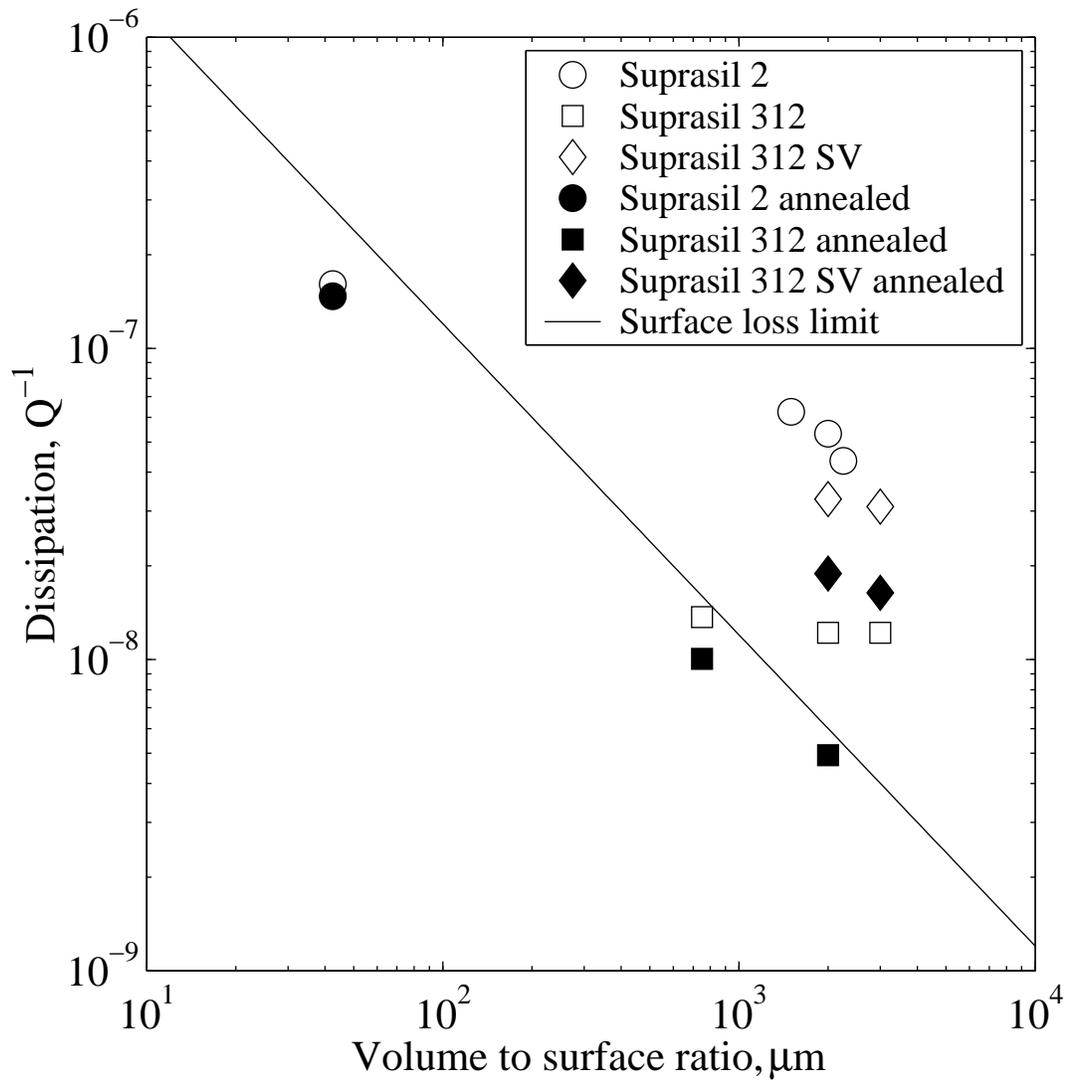}
\caption{The dependance of dissipation on volume to surface ratio.}
\label{graph}
\end{figure}

Table 1 lists the minimum measured loss level for each sample before and after annealing.
These loss levels are plotted against the samples' volume-to-surface ratio in Fig.~\ref{graph}. The line drawn in the figure was taken from the phenomenological model 
of Gretarsson and Harry~\cite{Andri} for Suprasil~2 fused silica fibers with diameter less than about 1 mm.
The line shows the scaling of dissipation expected if it were limited by surface losses, 
without significant levels of bulk losses in the material. 
Every data point in the figure is the smallest dissipation measured for each sample; 
usually this is seen in the second resonant mode. 

We have not fitted the thermoelastic losses for our samples, because  
thermoelastic losses are negligibly low for samples of the diameters 
we used in the flexural modes that we measured.

The highest quality factor was obtained for an annealed sample~8 mm in diameter produced from Suprasil~312. 
Before annealing the sample had the best quality factor  ($Q=82\times10^6$) of any unannealed sample.
This sample was accidentally broken and then it was fixed (we lost about 5 cm length at the end of the sample). 
The best $Q$ after the repair was $40\times10^6$. 
The best value for the annealed sample was $Q=2.03\times10^8$ 
(the loss angle $ \phi = 0.49\times 10^{-8}$) at a resonant frequency of 384~Hz.

Note that we obtained rather different results with different brands of fused silica. 
All of our best results were obtained with Suprasil~312. Samples of 312~SV showed 
substantially improved $Q$ after annealing, but even then the quality factors were 
poorer than unannealed Suprasil~312.

We found the strongest effects from annealing in our thickest  samples.
Our 8 and 12~mm samples showed $Q$ improvement of order a factor of 2, 
while our thinnest sample showed only a small improvement. 
This last observation is similar to what was seen by the Moscow group ~\cite{Mitr}.


\section{Discussion}

We can compare our results with the $Q$ measured by Numata and 
collaborators for modes in the range 30 to 100~kHz on a fused silica cylinder sample~\cite{Jap}.  
These results for Suprasil~312 show frequency-dependent $Q$s, 
with $Q$ decreasing with increasing frequency. (Comparison of the pattern of $Q$s to the mode shapes 
allowed a model to be made of contributions
from bulk losses, surface losses, and suspension losses.) Notably, the $Q$ values improved 
after annealing. Samples were annealed in a vacuum furnace 
for 24~hours at 980~$^\circ$C. 
The best $Q$ from these experiment was $34\times10^6$ before annealing (improving after annealing to $43\times10^6$), 
at a frequency of about 33~kHz. 
Our $Q$ at lower frequencies is much higher, but fits with the frequency dependence 
seen in the Numata experiments. 
Significant differences in values of the quality factors for samples from different material were noted
in Numata's work.

  Willems~\cite{Willems} has measured the quality factor of many modes of large 
unannealed samples of Suprasil 312 and 311SV. The gravest mode of the
312 sample had a $Q$ of 120 million (at a frequency of 11 kHz), about 50\% better 
than the best $Q$ seen in our unannealed samples.
Other modes had poorer $Q$, but there were many with $Q$ within a factor of 4 of the best. 
The best $Q$ of the 311SV sample was 50 million, 
again about 50\% better than the best $Q$ seen in our unannealed
311SV samples. The higher frequency modes of the two samples had quite comparable quality factors. 
Willems argues that this last fact demonstrates that the two samples have, in fact, 
similar dissipation levels overall, but it is also possible that the gap in best $Q$ is consistent with 
differences between 312 and 312 SV like those seen in our own work. Further
study will be required to untangle the comparison between 312 and 312SV.

In Fig.~\ref{graph} we drew the phenomenological law from Gretarsson and Harry~\cite{Andri} 
for comparison with the results of our measurements. Note that very few of our data points fall close to 
the line. (Perhaps this is because our data set is rather heterogeneous; it includes measurement 
of annealed and unannealed samples of three different brands of silica.)
Nevertheless, one can clearly see a trend to higher $Q$ at larger diameters, 
in qualitative agreement with the prediction that losses are dominated by surface effects, 
as opposed to dissipation in the bulk.

Our best result, $Q=2.03\times10^8$, shows that the quality factor of fused silica after annealing 
has reached a value comparable to that of sapphire at room temperature~\cite{VBthermal,Willems,Rowan}. 
This has led to renewed consideration of fused silica as a possible test mass material 
for Advanced LIGO\cite{Harryprivate}. 
However, it must be noted that Advanced LIGO would require silica with 
the low optical absorbtion of Suprasil 312~SV, where mechanical dissipation is not quite as good. 
Research on this question is in progress~\cite{Willems}.

\section{Acknowledgment}

We thank John Chabot, the glassblower at Syracuse University, for his skill and care in drawing the fused silica samples. For useful discussions, we thank Phil Willems, Kenji Numata, Gregg Harry, and Riccardo DeSalvo.
This work was support by Syracuse University and by National Science Foundation Grant No.~PHY-0140335


\begin{thebibliography}{30}

\bibitem{LIGO}
	A. Abramovici, et al., Science 265 (1992) 325.

\bibitem{VIRGO}
	A. Giazotto, et al., Nucl. Instrum. Methods A 289 (1988) 518.

\bibitem{GEO}
	K. Danzmann, et al., GEO 600: Proposal for a 600 m laser interferometric
	gravitatilnal wave antenna, Max-Planck-Institut fur Quantenptik Report 190, 
	Garching Germany, 1994.

\bibitem{TAMA}
	K. Tsubono, in: Gravitational Wave Experiments, Proc. First Edoardo Amaldi Conference, 
	Vol. 112, World Scientific, 1995.

\bibitem{Zener}
	C. Zener. Elasticity and anelasticity of metals, 
	Univ. of Chicago Press, Chicago, 1948.

\bibitem{VBthermal}
	V.B. Braginsky, V.P. Mitrofanov, V.I. Panov, 
	Systems with small dissipation, Univ. of Chicago Press, Chicago, 1985.   

\bibitem{Andri}
	A. Gretarsson, G. Harry, Rev.Sci. Instrum. 70 (1999) 4081.

\bibitem{Mitr}
	V.P. Mitrofanov, K.V. Tokmakov, Phys. Lett. A 308 (2003) 212.

\bibitem{Bartenev}
Bartenev, G. M., Lomovskoi, V. A., and Sinitsyna, G. M., {\em Inorganic
Materials}, {\bf 32} (1996) p. 671-682

\bibitem{Lunin}
Lunin, B. S., and Torbin, S. N., {\em Moscow University Chemistry Bulletin}
October 2002

\bibitem{Zdaniewski}
Zdaniewski, W. A., Rindone, G. E., Day, D. E., {\em Journal of Material
Science}, {\bf 14} (1979) 763-775.

\bibitem{Fraser}
Fraser, D. B., {\em Journal of Applied Physics}, {\bf 41} (1970) 6-11.


\bibitem{Jap}
	K. Numata, S. Otsuka, M. Ando and K. Tsubono, Class. Quantum Grav. 19 (2002) 1697.
	
\bibitem{Lunin}
      B. Lunin, 2003, Moscow State University. Private communication.

\bibitem{Parc}
	R. Le Parc, B. Champagnon, Ph. Guenot and S. Dubois., J. Non-Cryst. Solids 293-295 (2001) 366.

\bibitem{Sakaguchi}
	S. Sakaguchi, S. Todoroki and T. Murata., J. Non-Cryst. Solids 220 (1997) 178.

\bibitem{Steve}
	S. Penn, G. Harry, A. Gretarsson, S. Kittelberger, P. Saulson, J. Schiller, 
      J. Smith, S. Swords, Rev. Sci. Instrum. 72 (2001) 3670.

\bibitem{Willems}
	P. Willems, 2003, California Institute of Technology. Private communication.	

\bibitem{Rowan}
	S. Rowan, G Cagnoli, P. Sneddon, J. Hough, R. Route, E.K. Gustafson, M.M. Fejer and V.P. Mitrifanov.
	 Phys. Lett. A 265 (2000) 5. 

\bibitem{Harryprivate} 
	G. Harry, 2003, Massachusetts Institute of Technology. Private communication.



\end{thebibliography}
\end{document}